\documentstyle[aps,preprint,prb]{revtex}
\begin{document}
\def\vk{\vec k} 
\def\br{{\bf r}}
\title{\bf Absence of the Vortex Solution in Gor'kov's Formalism}
\author{Yong-Jihn Kim }
\address{Department of Physics,  Bilkent University,\\
 06533 Bilkent, Ankara, Turkey}
\maketitle
\begin{abstract}

It is shown that the Abrikosov's vortex solution or its corresponding 
two-particle pair potential is not the solution of
the self-consistency equation in Gor'kov's formalism.
Since the self-consistency equation leads to a superposition of 
different types of off-diagonal long-range order (ODLRO) instead of 
one type of ODLRO only, it may not handle
the vortex problem appropriately. A possible resolution is suggested. 

\end{abstract}

\vskip 5pc
PACS numbers: 74.50.+r, 74.60.-w, 74.60.Ec

\vspace{1pc}

\noindent

\vfill\eject
\section{\bf Introduction} 

Recent STM experiments$^{1-3}$ show that the microscopic structures
of the vortex state in type II and high $T_{c}$ superconductors
are much more complicated than expected. In particular, the vortex core and
the lattice structure depend strongly on the microscopic details of the samples.
For instance, using $NbSe_{2}$ samples Hess and his collaborators$^{1,2}$ 
found unexpected
zero-bias peaks and sixfold star-shaped structures in the vortex core states. 
These experiments revived much theoretical interest.$^{4-6}$ However, a 
satisfactory quantitative explanation is still not available.$^{7}$
Maggio-Aprile et al.$^{3}$ also reported that the vortices in 
$YBa_{2}Cu_{3}O_{7-\delta}$ arrange in an oblique lattice, which 
remains unexplained.  On theoretical side, Thouless et al.$^{8}$ noted that 
there is still a lot of controversy in the dynamics of vortices.

It seems that there is some fundamental problem in our theoretical
understanding of the vortex state in superconductors. In this paper,
we point out that Gor'kov's Green's function formalism or the 
Bogoliubov-de Gennes equations may not handle the vortex problem properly, 
since the self-consistency equation gives rise to an incoherent superposition
of different types of off-diagonal long-range-order (ODLRO) instead of 
a coherent vortex state. 
The same problem was already found in Gor'kov and Galitskii's  (GG)$^{9}$ 
solution for the d-wave pairing BCS theory. GG allowed a superposition 
of several distinct types of the off-diagonal long-range-order, which was 
proven to be invalid.$^{10-14}$
As Anderson noted,$^{13}$ this problem occurs in many cases even in
discussing flux quantization. 
In fact, it also hindered correct understanding of the effects of magnetic 
impurities$^{15,16}$ and weak localization$^{17-20}$ on superconductivity.
Balian, Nosanow, and Werthamer$^{11}$ showed 
that this problem is caused by the difficulty in the method of 
Green's functions for a manybody system. It is inspiring to remind their
statement: {\sl Thus the Green's function method as usually formulated is not a 
complete dynamical description of the system, and requires in addition, 
some criterion to distinguish these extraneous solutions from the correct one}.
The criterion may be provided by the physical constraint of the Anomalous 
Green's function, which is nothing but the pairing constraint.$^{17,18}$

The vortex problem may be very complicated; so we use the following 
simplifications which are not crucial in our discussion though.

(i) We consider only the lowest Landau Level in the presence of the very high 
magnetic fields, i.e.,
\begin{equation}
\omega_{c}>\omega_{D},
\end{equation}
where $\omega_{c}$ and $\omega_{D}$ are the cyclotron frequency and 
the Debye frequency, respectively.

(ii) The Zeeman splitting is disregarded.

(iii) We use the Landau gauge where 
\begin{equation}
A_{x}=0, A_{y}=Bx.
\end{equation}
The z-axis motion will be suppressed.

(iv) We consider the self-consistency equation near $T_{c}$.

\section{Landau Levels and Abrikosov's Vortex Solution} 

We consider a rectangular sample with sides $L_{x}$ and $L_{y}$.
For the Hamiltonian
\begin{equation}
H= {p_{y}^{2}\over 2m} + {(p_{x}-{eBx\over c})^{2}\over 2m},
\end{equation}
the eigenfunctions are given by
\begin{equation}
\phi_{n}(x,y)=N_{o}e^{iqny}exp[-(x-qn\ell^{2})^{2}/2\ell^{2}],
\end{equation}
where 
\begin{equation}
\ell=\sqrt{\hbar c/eB},
\end{equation}
and 
\begin{equation}
 q={2\pi\over L_{y}}.
\end{equation}
Here $N_{o}$ is a normalization constant and 
we have used the periodic boundary condition along the y-direction.
If the x dimensions of the system are confined to $-L_{x}/2<x<L_{x}/2$,
n is determined by the condition
\begin{equation}
-{L_{x}\over 2} < nq\ell^{2} < {L_{x}\over 2}.
\end{equation} 
(We neglect the effect of boundary on $N_{o}$.)

The famous Abrikosov's vortex solution is$^{21}$ 
\begin{equation}
\Psi(x,y)=\sum_{n=-\infty}^{\infty}C_{n}
e^{iqny}exp[-(x-qn\ell^{2})^{2}/2\ell^{2}],
\end{equation}
with the periodicity conditions, $C_{n+1} =C_{n}$ for a square
lattice and $C_{n+2}=C_{n}$ for a triangular lattice,$^{22}$ respectively.
The $C_{n}$ and $\ell$ are adjusted to minimize the free energy in the 
Ginzburg-Landau theory.  Note that the solution is a linear combination of 
one-particle eigenfunctions.
  
\section{\bf Self-consistency Equation} 

In terms of the Lowest Landau Level wavefunctions, the normal state 
Green's function $G$ is written as
\begin{equation}
G_{\omega}(1,2)=\sum_{n}{\phi_{n}(x_{1},y_{1})\phi^{*}_{n}(x_{2},y_{2})\over i\omega -\epsilon_{o}},
\end{equation}
where $\epsilon_{o}={1\over 2}\hbar \omega_{c}-\mu$.$^{23}$ $\mu$ is the Fermi 
energy and $\omega$ are the Matsubara frequencies.
The Gor'kov's self-consistency equation is then given by$^{24}$ 
\begin{equation}
\Delta(x_{1},y_{1})=VT\int\sum_{\omega}G^{\uparrow}_{\omega}(1,2)
G_{-\omega}^{\downarrow}(1,2)\Delta(x_{2},y_{2})dx_{2}dy_{2}.
\end{equation}

According to the Gor'kov's microscopic derivation of the Ginzburg-Landau 
equations,$^{24}$ the order parameter $\Psi({\bf r})$ is proportional to 
the (two-body) pair potential $\Delta({\bf r})$:
\begin{equation}
\Psi({\bf r})=\Delta({\bf r})\sqrt{7\zeta(3)N}4\pi T_{c},
\end{equation}
where $\zeta$ is Riemann's zeta function and N is the electron number density
in the normal metal.
Consequently, the Abrikosov's vortex solution Eq. (8) should be
the solution of the above self-consistency equation. 
Unfortunately, this is not the case.
For example, if we substitute the pair potential
\begin{equation}
\Delta(x,y)\sim C_{1}e^{iqy}exp[-(x-q\ell^{2})^{2}/2\ell^{2} ], 
\end{equation}
into the self-consistency equation, we obtain the different form of the
pair potential 
\begin{equation}
\Delta(x,y)\sim C_{1}e^{iqy}exp[-(x-q\ell^{2}/2)^{2}/\ell^{2}]. 
\end{equation}
In other words, the self-consistency condition is not satisfied.
This evidence cast serious doubt on the Gor'kov's microscopic derivation of
the Ginzburg-Landau equations.   
In fact, this difficulty is anticipated since the Abrikosov's vortex solution
is a linear combination of one-particle eigenfunctions while the 
pair potential consists of the multiplication of two one-particle 
eigenfunctions.

\section{\bf Absence of the Vortex Solution } 

Another possible form of the pair potential is
\begin{equation}
\Delta(x,y)=\sum_{n} \Delta_{n}e^{iqny}exp[-(x-qn\ell^{2}/2)^{2}/\ell^{2}]. 
\end{equation}
Now we show that this pair potential is not the solution of the 
self-consistency equation, either.
First, we consider the constant pair potential in the y-direction, i.e.,
\begin{equation}
\Delta(x,y)\sim  \Delta_{0}exp(-x^{2}/\ell^{2}). 
\end{equation}
Upon substitution of this simple solution into Eq. (10), we find pairing
between $\phi_{n}\uparrow$ and $\phi_{-n}\downarrow$ and the transition
temperature $T_{c}$ is determined by
\begin{equation}
1=VT_{c}{N_{o}^{2}\over \sqrt{2}} \sum_{\omega}\sum_{n} {e^{-2n^{2}q^{2}\ell^{2}}\over \omega^{2}+\epsilon_{o}^{2}}.
\end{equation}
We could have obtained the same equation from the BCS theory with the pairing 
matrix elements, 
\begin{eqnarray}
V_{nn'} &=& V\int \phi_{n}^{*}(r) \phi_{-n}^{*}(r) \phi_{-n'}(r) \phi_{n'}(r)
                  d{r}\nonumber\\
&=& V{N_{o}^{2}\over \sqrt{2}}e^{-(n^{2}+n'^{2})q^{2}\ell^{2}}.
\end{eqnarray}
The manybody ground state is, then 
\begin{equation}
\tilde{\phi}_{BCS}=\prod_{n}[u_{n}+v_{n}(\phi_{n}\uparrow, \phi_{-n}\downarrow)]
|0>
\end{equation}

Second, we consider the pair potential for $n=1$ in Eq. (14),
\begin{equation}
\Delta(x,y)\sim \Delta_{1}e^{iqy}exp[-(x-q\ell^{2}/2)^{2}/\ell^{2}]. 
\end{equation}
It is straightforward to show that this pair potential leads to pairing between
$\phi_{n}\uparrow$ and $\phi_{-n+1}\downarrow$ and the resulting transition
temperature $T_{c}'$ is determined by
\begin{equation}
1=VT_{c}'{N_{o}^{2}\over \sqrt{2}} \sum_{\omega}\sum_{n} 
{e^{-2(n-1/2)^{2}q^{2}\ell^{2}}\over \omega^{2}+\epsilon_{o}^{2}}.
\end{equation}
In the BCS theory, the corresponding pairing matrix elements are 
\begin{eqnarray}
V_{nn'} &=& V\int \phi_{n}^{*}(r) \phi_{-n+1}^{*}(r) \phi_{-n'+1}(r) \phi_{n'}(r)
                  d{r}\nonumber\\
&=& V{N_{o}^{2}\over \sqrt{2}}e^{-(n-1/2)^{2}q^{2}\ell^{2}}
e^{(n'-1/2)^{2}q^{2}\ell^{2}}.
\end{eqnarray}
Note that the transition temperatures may be extremely small due to the 
exponential factors and $T_{c}$ and $T_{c}'$ are different.
The ground state is now
\begin{equation}
\tilde{\phi}_{BCS}'=\prod_{n}[u_{n}+v_{n}(\phi_{n}\uparrow, \phi_{-n+1}\downarrow)]
|0>.
\end{equation}

If we combine the two solutions, we obtain the pair potential
 \begin{equation}
\Delta(x,y)\sim  \Delta_{0}exp(-x^{2}/\ell^{2}) 
+ \Delta_{1}e^{iqy}exp[-(x-q\ell^{2}/2)^{2}/\ell^{2}]. 
\end{equation}
Inserting Eq. (23) into Eq. (10) one finds
 \begin{equation}
 \Delta_{0}exp(-x_{1}^{2}/\ell^{2}) 
=VT\int\sum_{\omega}G^{\uparrow}_{\omega}(1,2)
G_{-\omega}^{\downarrow}(1,2)\Delta_{0}exp(-x_{2}^{2}/\ell^{2}) 
dx_{2}dy_{2},
\end{equation}
and 
 \begin{eqnarray}
 \Delta_{1}e^{iqy_{1}}exp[-(x_{1}-q\ell^{2}/2)^{2}/\ell^{2}] 
&=&VT\int\sum_{\omega}G^{\uparrow}_{\omega}(1,2)
G_{-\omega}^{\downarrow}(1,2)\nonumber\\
&\times& \Delta_{1}e^{iqy_{2}}exp[-(x_{2}-q\ell^{2}/2)^{2}/\ell^{2}] 
dx_{2}dy_{2},
\end{eqnarray}
since the two solutions are linearly independent.
Notice that we can not find the temperature at which the two different types
of the condensation occur simultaneously due to the difference in
the pairing matrix elements as shown above.
Nevertheless, as Galitskii$^{25}$ suggested, it is tempting to write the 
resulting manybody state as a 
combination of the above ground states, that is,
 \begin{equation}
\tilde{\phi}_{BCS}+ \tilde{\phi}_{BCS}'=
\prod_{n}[u_{n}+v_{n}(\phi_{n}\uparrow, \phi_{-n}\downarrow)]|0>
+
\prod_{n}[u_{n}+v_{n}(\phi_{n}\uparrow, \phi_{-n+1}\downarrow)]|0>.
\end{equation}
This combination is just the so-called incoherent superposition of 
different types of off-diagonal long-range-order(ODLRO),$^{9-14}$
which does not correspond to any real physical state.
Hone$^{10}$ actually demonstrated the impossibility of constructing a complete
hierarchy of Green's functions in such a case.

For the pair potential 
\begin{equation}
\Delta(x,y)\sim \Delta_{2}e^{i2qy}exp[-(x-q\ell^{2})^{2}/\ell^{2}], 
\end{equation}
one finds the $T_{c}$ equation
\begin{equation}
1=VT_{c}{N_{o}^{2}\over \sqrt{2}} \sum_{\omega}\sum_{n} 
{e^{-2(n-1)^{2}q^{2}\ell^{2}}\over \omega^{2}+\epsilon_{o}^{2}}.
\end{equation}
When the electrons are confined in the x-direction as assumed here, 
the $T_{c}$ is different from those of the previous cases.
Whereas for an infinite system or the system with the periodic boundary 
condition in the x-direction, the $T_{c}$ may be the same as that for the pair 
potential corresponding to $n=0$ in Eq. (14).
We have then $T_{c}$ for $\Delta_{0}, \Delta_{2}, \Delta_{4}, \cdots$ and 
$T_{c}'(\not=T_{c})$ for $\Delta_{1}, \Delta_{3}, \Delta_{5}, \cdots$. 
It is interesting to note that the Abrikosov solution with $C_{n+2}=C_{n}$ has
a lower energy than that with $C_{n+1}=C_{n}$.

Thus, the self-consistency equation does not allow the
Abrikosov's vortex solution or the corresponding two-particle pair
potential Eq. (14). It is obvious that adding different types of off-diagonal
long-range-order (ODLRO) in Gor'kov's formalism does not lead to a coherent
vortex state.
On the other hand, previous workers first sum over the eigenstate $\phi_{n}$
in the one-particle Green's function and then consider the self-consistency
condition, which fails to take into account the two-particle correlations 
correctly. 

\section{\bf Discussion } 

To describe the vortex state, we need to devise a coherent manybody ground
state which may be closely related to the Abrikosov's vortex solution. 
Feynman's vortex solution of the superfluid He-4 may be a good starting point.
More details will be published elsewhere.$^{26}$ 
Gor'kov's formalism may also be generalized to obtain a coherent superposition
of the different types of the pairing.$^{17,18}$ 

\vspace{1pc}
\centerline{\bf ACKNOWLEDGMENTS}

I am grateful to Hatice Altug for discussions about Feynman's wavefunction.


\begin{references}

\bibitem{1} H. F. Hess, R. B. Robinson, R. C. Dynes, J. M. Valles, Jr., and J. V. Waszczak, Phys. Rev. Lett. {\bf 62}, 214 (1989). 
\bibitem{2} H. F. Hess, R. B. Robinson, and J. V. Waszczak, Phys. Rev. Lett. {\bf 64}, 2711 (1990). 
\bibitem{3} I. Maggio-Aprile, Ch. Renner, A. Erb, E. Walker, and O. Fischer, Phys. Rev. Lett. {\bf 75}, 2754(1995). 
\bibitem{4} A. W. Overhauser and L. L. Daemen, Phys. Rev. Lett. {\bf 62}, 1691 (1989). 
\bibitem{5} J. D. Shore, M. Huang, A. T. Dorsey, and J. P. Sethna, Phys. Rev. Lett. {\bf 62}, 3089 (1989). 
\bibitem{6} F. Gygi and M. Schluter, Phys. Rev. Lett. {\bf 65}, 1820 (1990). 
\bibitem{7} Harald Hess's web page.
\bibitem{8} D. J. Thouless, P. Ao, Q. Niu, M. R. Geller, and C. Wexler, cond-mat/9709127.
\bibitem{9} L. P. Gor'kov and V. M. Galitskii, Sov. Phys. JETP. {\bf 13}, 792 (1961).
\bibitem{10} D. Hone, Phys. Rev. Lett. {\bf 8}, 370 (1962).
\bibitem{11} R. Balian, L. H. Nosanow, and N. R. Werthamer, Phys. Rev. B {\bf 8}, 372 (1962).
\bibitem{12} P. W. Anderson, Bull. Am. Phys. Soc. {\bf 7}, 465 (1965).
\bibitem{13} P. W. Anderson, Rev. Mod. Phys. {\bf 38}, 298 (1966).
\bibitem{14} P. W. Anderson, {\sl Basic Notions of Condensed Matter Physics}, (Benjamin/Cummings, Menlo Park, 1984), p. 247.
\bibitem{15} Yong-Jihn Kim and A. W. Overhauser, Phys. Rev. B {\bf 49}, 15799 (1994).
\bibitem{16} Mi-Ae Park, M. H. Lee, and Yong-Jihn Kim, Physica C {\bf 306}, 96 (1998). 
\bibitem{17} Yong-Jihn Kim, Mod. Phys. Lett. B {\bf 10}, 555 (1996).
\bibitem{18} Yong-Jihn Kim, Int. J.  Mod. Phys. B {\bf 11}, 1731 (1997).
\bibitem{19} Yong-Jihn Kim and K. J. Chang, Mod. Phys. Lett. B {\bf 12}, 763 (1998).
\bibitem{20} Mi-Ae Park and Yong-Jihn Kim, cond-mat/9909365.
\bibitem{21} A. A. Abrikosov, Zh. Eksperim. i Teor. Fiz. {\bf 32}, 1442 (1957)
[Sov. Phys. JETP {\bf 5}, 1174 (1957)]. 
\bibitem{22} W. H. Kleiner, L. M. Roth, and S. H. Autler, Phys. Rev. {\bf 133}, A1226 (1964).
\bibitem{23} Since this quantity becomes zero for an ideal 2-d problem, we 
actually need to include the motion along the z-direction, which can be done 
easily. Alternatively we may include the edge state. See B. I. Halperin, Phys. Rev. 
B {\bf 25}, 2185 (1982).  This does not affect our conclusion. 
\bibitem{24} L. P. Gor'kov, J. Exptl. Theor. Phys. (U.S.S.R) {\bf 36}, 1918 (1959) [Sov. Phys. JETP {\bf 36}, 1364 (1959)].
\bibitem{25} V. M. Galitskii, Physica {\bf 26}, S143 (1960).
\bibitem{26} Hatice Altug and Yong-Jihn Kim, unpublished.

\end{references}
\end{document}